\documentstyle[12pt,epsf]{article}
\newcommand{\cc}{\cite}

\newcommand{\beq}{\begin{equation}}
\newcommand{\eeq}{\end{equation}}
\newcommand{\ppd}{\partial}
\newcommand{\epps}{\epsilon}
\newcommand{\vepps}{\varepsilon}

\title{A finite-energy solution in Yang-Mills theory and quantum
       fluctuations.}

\author{ O.V. Pavlovski\u{\i} \thanks{e-mail address:
ovp@goa.bog.msu.su }  \\ {\em Bogoliubov Institute for Theoretical
Problem of Microphysics,} \\ {\em Lomonosov Moscow State University
(MSU) } \\ {\em Moscow, 119899, Russian Federation.} }

\date{ \ \ \  }

\begin{document}

\maketitle

\begin{abstract}
A finite-energy solution of Yang-Mills theory with a nonstandard
lagrangian is provided.  Properties of these solution are studied
and also a possible physical interpretation is given.
\end{abstract}

\vspace{1cm}

PACS number(s): 11.10.Lm, 11.27.+d, 11.15.Tk, 12.39.Mk

\section{Introduction.}

This paper is devoted to constructing of the stable finite-energy
and compact gluon object in the Yang-Mills theory with a nonstandard
lagrangian. This lagrangian consists of pure Yang-Mills lagrangian
${\cal L}_{YM} = -1/4 \, (F_{\mu \nu})^a (F^{\mu \nu})^a$ and higher
derivative term $\Delta {\cal L}$ related to quantum
fluctuations of the gluon field. Such an approach to investigation of
quantum fluctuations had been introduced in the series of papers in
the middle of 80's \cc{e1}-\cc{ale2}. In these
papers IR low-energy limit of QCD have been studied and there was
demonstrated that contribution from the quantum fluctuations can be
taken into account by modification of QCD lagrangian via introducing
into the lagrangian the additional higher derivative terms like
$\epps^{abc} (F_{\mu \nu})^a ({F^\nu}_\rho)^b (F^{\rho \mu})^c$ or
$(D_\rho F_{\mu \nu})^a (D^\rho F^{\mu \nu})^a$. From methodological
point of view lagrangians obtained within this approach are very
similar to the well-known Euler-Heisenberg effective lagrangian in
QED \cc{EH}. As a result, in leading approximation for gluon field one
obtained the theory with a new lagrangian ${\cal L}={\cal
L}_{YM}+\Delta {\cal L}$ for the $c$-number field $A_{eff}$ and this
classical Yang-Mills field is an average of the initial gluon field
$A_0$ over quantum fluctuations:  $A_{eff}=\langle A_0 \rangle$. In
\cc{e3,e4,e5} the investigation of classical solutions in such
effective theories was discussed in context of the color confinement
problem. A very similar problem is discussed in paper \cc{Dir}
also.

There are many approaches to obtaining  finite-energy compact gluon
objects from QCD. The crucial point here stems from the fact that
pure classical Yang-Mills theory hasn't such solutions by  reason of
scale invariance. In the papers \cc{lun_b,lun_pav} it was shown that
typical spherically symmetrical solution of pure Yang-Mills theory is
an infinite-energy solution with singularity on the finite radius
sphere. Therefore, any approaches to finding gluon clusters is
based on the various modifications of lagrangians. For example, the
well-known monopole solution \cc{mon} exists thanks to the
interaction of Yang-Mills field with the field of matter.
Furthermore, some kind of gluon finite-energy objects appear in
lattice approach to the QCD \cc{gl-ball}. Such solutions are called
glueballs. A glueball on the lattice is a quantum object having
no analogues in classical field theory. Unfortunately such colorless
gluon object wasn't be observed by experiment yet and only model
predictions for physical characteristics like mass, effective radius
and so on exist now.

In the present work we try to find similar quantum compact
finite-energy objects by using of an effective low-energy approach to
the QCD that was discussed above.

\section{Finite-energy gluon clusters.}

In this section we investigate the classical Yang-Mills theory with
a nonstandard modified lagrangian
\beq
{\cal L}^\vepps_{YM} = -\frac{1}{4}(F_{\mu \nu})^a (F^{\mu
\nu})^a -\frac{\vepps^2}{6} \epps^{abc} (F_{\mu \nu})^a
({F^\nu}_\rho)^b (F^{\rho \mu})^c
\label{2.1}
\eeq
where $\vepps=1/M$ is an inverse mass dimensional parameter
characterizing the intensity of quantum fluctuation;
$(F_{\mu \nu})^a=\ppd_\mu A^a_\nu - \ppd_\nu A^a_\mu +
\epps^{abc} A^b_\mu A^c_\nu $ and
$(D_\mu)^{ab}=\delta^{ab} \ppd_\mu + \epps^{abc} A^c_\mu$.
Here we deal with $SU(2)$ Yang-Mills field.

Such form of modification of the Yang-Mills lagrangian is chosen due
to the fact that the theory obtained  contains only second-order
derivative terms. Thus the dynamics of this field theory can be
studied in detail.

Using the variation principle, we get the equation of motion
\beq
D^{ab}_\mu (F^{\mu \nu} - \vepps^2 G^{\mu \nu})^b=0.
\label{2.2}
\eeq
where $(G^{\mu \nu})^a=\epps^{abc}({F^\nu}_\rho)^b (F^{\rho
\mu})^c$.

Adding the divergence
\beq
\ppd_\rho [ (F^{\nu \rho}-
\vepps^2 G^{\nu \rho})^a \ppd_\mu A_\rho^a ],
\label{2.4}
\eeq
to the energy-momentum tensor
\beq
 {T^\nu}_\mu=  \ppd_\mu A^a_\rho
 \frac{ \ppd  {\cal L}^\vepps_{YM}}{\ppd(\ppd_\nu A^a_\rho)} -
{\delta^\nu}_\mu  {\cal L}^\vepps_{YM} =
-(F^{\nu \rho}- \vepps^2 G^{\nu \rho})^a \ppd_\mu A_\rho^a
-{\delta^\nu}_\mu {\cal L}^\vepps_{YM},
\label{2.3}
\eeq
we obtain the symmetrical form of this tensor
\beq
 {T^\nu}_\mu= -(F^{\nu \rho}- \vepps G^{\nu \rho})^a
(F_{\mu \rho})^a -{\delta^\nu}_\mu {\cal L}^\vepps_{YM} .
\label{2.5}
\eeq

Now we consider the spherically symmetrical chromomagnetic field
configuration. Substituting the well-known \cc{wy} Wu-Yang ansatz
\beq
A^a_0=0, \qquad A^a_i=\epps_{aij} n_j
\frac{1-H(r)}{r},
\qquad \qquad n_i=x_i/r \qquad r=\sqrt{x_i^2} ,
\label{2.6}
\eeq
in  (\ref{2.2}),  we get the
following equation on amplitude $H(r)$
$$ \left(1 -
\frac{\vepps^2}{r^2} (H(r)^2-1) \right) r^2 H(r)'' = H(r) \Bigl(
H(r)^2-1 \Bigr) + $$ \beq +\frac{\vepps^2}{r^2} \Bigl((rH(r)')^2 H(r)
- 2 r H(r)'(H(r)^2-1)\Bigr).
\label{2.10}
\eeq
Energy of field configuration generated by the solution of equation
(\ref{2.10})  $H(r)$ is the functional
\beq
E^\vepps [H] = \int T^{00} d^3 x =
4\pi \int\limits_0^\infty
\left[ \left(1 - \frac{\vepps^2}{r^2} (H(r)^2-1) \right)
\left(H(r)'\right)^2 +
\frac{(H(r)^2-1)^2}{2r^2} \right] dr =
\int\limits_0^\infty E(r) dr  .
\label{2.11}
\eeq

The next aim of our investigation is finding the solutions of
equation (\ref{2.10}). Notice that only finite-energy solutions are
interesting  for us. Hence the functional $E^\vepps [H]$ (\ref{2.11})
should be finite on such solutions.

Equation (\ref{2.10}) is a very complicated nonlinear differential
equation. In order to solve it only numerical or approximation
methods seem applicable. The crucial point of such analysis is that
the leading derivative term in this equation contains the factor
\beq
\Phi [H] (r) = \Bigl( r^2 - \vepps^2 (H(r)^2-1)  \Bigr).
\label{3.1}
\eeq
If $H_s(r)$ is a solution of equation (\ref{2.10}) and there is a
point $r=R$ such that $\Phi [H_s] (R)=0$, then this solution $H_s$
has singular behavior in a neighborhood of this point $r=R$ due to
smallness of the factor $\Phi [H_s]$. Using the standard
procedure, one obtains the asymptotic behavior near this point
\beq
H_s (r) \stackrel{\strut r \to R \pm 0} {\longrightarrow}   \pm
\sqrt{ 1+ R^2/\vepps^2 } - C \, (R-r)^{2/3} + \underline{O}(R-r),
\label{3.3}
\eeq
where $C$ is a constant. Of course, the $H_s(R)$
finite but its derivative at the point $r=R$ is singular. Indeed,
\beq
{H'}_s (r) \stackrel{\strut r \to R \pm 0} {\longrightarrow}
\frac{2}{3} C \, (R-r)^{-1/3} + \underline{O}(1) \longrightarrow
\infty.  \label{3.4}
\eeq

Such singular behavior is analogous to the  singular
behavior on finite sphere of solutions of pure Yang-Mills field
\cc{lun_pav} discussed above but there is a principal difference.
Energy of such
solutions in the pure Yang-Mills case is infinite but in the modified
Yang-Mills case energy (and other physical characteristics) of
solution with singular behavior (\ref{3.4}) should be {\it finite}:
\beq
E(r)|_{r=R} \sim  4 \pi \left( \pm \frac{8 \vepps^2}{9 R^2} \sqrt{ 1+
R^2/\vepps^2 } \, C^3 + \frac{R^2}{2 \vepps^4} \right) <\infty .
\label{3.5}
\eeq
Therefore such solutions are {\it physical}.

Now we should discuss the numerical investigation of solutions of
equation (\ref{2.10}) that have the asymptotic behavior (\ref{3.4})
at some point $r=R$.

To guarantee stability of our solutions we should choose this
asymptotics at origin ($r \to 0$)
\beq
H(r) \simeq  -1 + a_1\, r^2 +
 a^2_1\, \frac{2\vepps^2 a_1 - 3 }{10(1 + 2  a_1 \vepps^2)}\, r^4
+ \underline{O}(r^6),
\label{3.6}
\eeq
and at infinity ($r \to \infty$)
$$
H(\rho) \simeq  1 + a_2\, \rho + \frac{3}{4} a^2_2\, \rho^2 +
\frac{11}{20}  a^3_2\, \rho^3 +
\frac{193 a^2_2 -240 \vepps^2 }{480} a^2_2 \rho^4 +
$$
\beq
+\frac{329 a^2_2 -1280 \vepps^2 }{1120} a^3_2 \rho^5 +
\underline{O}(\rho^6), \qquad  \rho=1/r,
\label{3.7}
\eeq
where $a_1>0$ and $a_2>0$ are constants.
Solutions with such asymptotic are stable because vacuum states
$H(0)=-1$ and $H(\infty)=1$ are different.

Notice that equation (\ref{2.10}) has very useful symmetries.
First of all, this equation is symmetrical with respect to the
changes $H \leftrightarrow -H$. So, if we have a solution $H(r)$,
then $-H(r)$ is a solution too.

\begin{figure}[t]
\leavevmode
\epsfbox{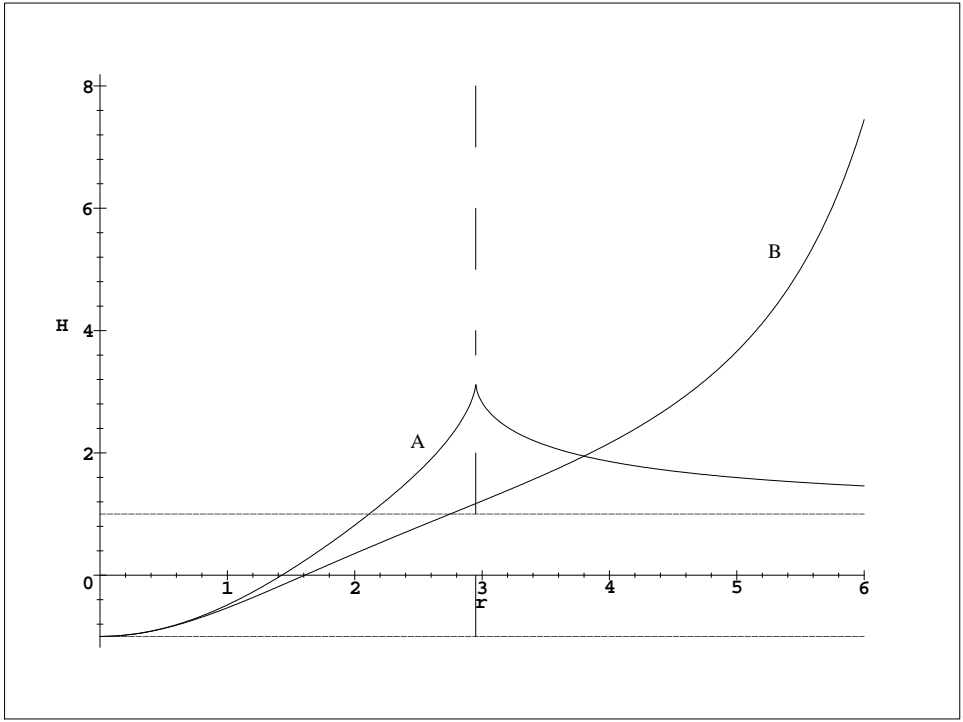}
\caption{ Gluon cluster object. Amplitude $H(r)$ ($\vepps=1$,
$a_1=0.544$ and  $R=1.425$). }
\end{figure}

\begin{figure}[t]
\leavevmode
\epsfbox{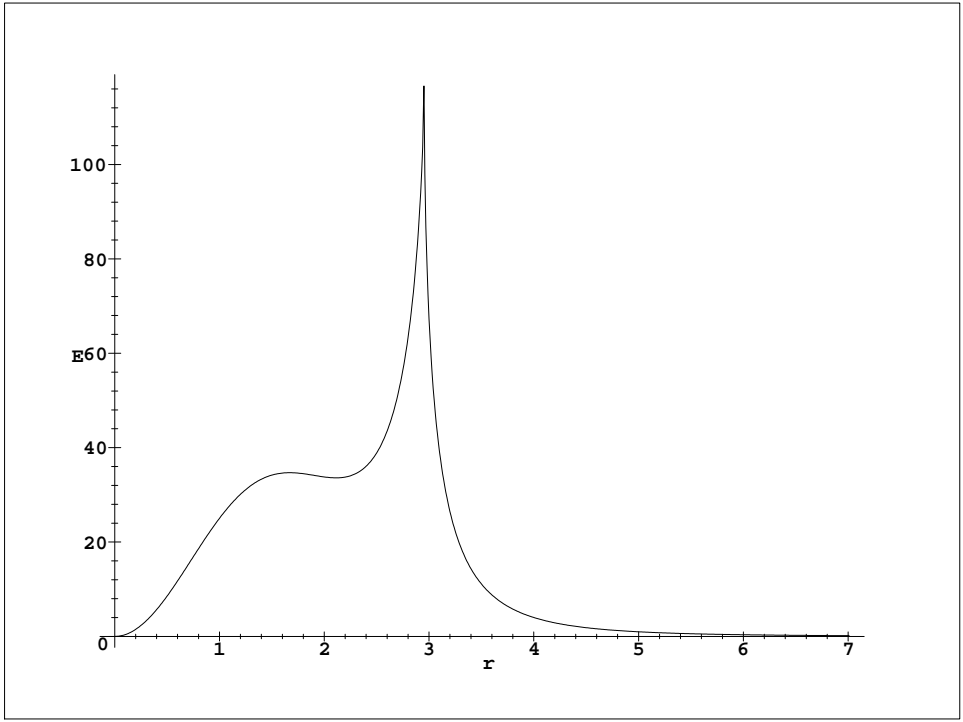}
\caption{ Gluon cluster object. Density of energy.}
\end{figure}

Now let $\vepps=\vepps_1$ and we have a set  of
solutions $\{H_{\vepps_1}(r)\}$.  If we perform the change of
variable
\beq
r \to \frac{\vepps_1}{\vepps_2} r,  \qquad
\{H_{\vepps_1}(r)\}
\stackrel{\strut r \to \vepps_1 r/\vepps_2}{\Longrightarrow}
\{H_{\vepps_2}(r)=H_{\vepps_1}(\frac{\vepps_1}{\vepps_2} r) \},
\label{3.8}
\eeq
we get equation (\ref{2.10}) again but with new
$\vepps=\vepps_2$. If we know a solution for some $\vepps>0$, say,
$\vepps=1$, then using (\ref{3.8}) we can obtain a solution for
any other $\vepps_1>0$.

The numerical investigation of equation (\ref{2.10}) is
presented in Fig.1. The solution of pure Yang-Mills theory
($\vepps=0$) is shown by the line B. The function A is the solution
of equation (\ref{2.10}) with $\vepps=1$. In Fig.2 we can see the
energy density (\ref{2.11})
corresponding to this solution.

The solution $H_<$ starting at the origin (or {\it internal})
increases monotonically and its energy density increases too.
Evidently, as energy density grows, the role of quantum fluctuations
grows too. At the point $r=R$, the energy density attains its
critical value $E^{cr}$  and H(r) becomes singular (\ref{3.3}).
The solution $H_>$ starting at the infinity (or {\it external})
demonstrates an absolutely similar behavior. Now, a vary essential
questions arises: How to connect these two sets of solutions and
how to determine such solutions on the whole space?

These questions have no mathematical answer because in this case we
deal with the solutions that can not be extended to the right (to the
left) because the point of singularity $r=R$ is essential.

Obviously, this nonuniqueness of solution in the whole space is due
to underdetermination of our effective model. It is necessary to
introduce some additional physical condition that would allow to
choose a physically reasonable solution from the broad class of
solutions described above.

Since this solution of the model (\ref{2.1}) has to be an effective
approximation to an existing gluon object, the general properties of
the latter should be  represented by the former. Thus, if energy
density of this gluon object is continuous everywhere, then it should
be continuous for the approximating solution of equation (\ref{2.10})
as well. We show below that the condition of continuity of energy
density is sufficient for the construction of a unique solution and
investigation of its properties.

According to mathematical structure of solutions of this model, the
condition of continuity of energy density can be formulated as
follows: {\it There exists a critical density of the
energy for classical solutions in our effective Yang-Mills theory
and the value of this critical density $E^{cr}$
is a  physical property of the theory. Therefore $E^{cr}$ shouldn't
depend on kind of solution (internal or external).} It follows that
\beq
E(r)_<|_{r=R}=E(r)_>|_{r=R} \qquad {\Longrightarrow} \qquad C_<=C_>.
\label{3.10} \eeq

It is easily shown that condition (\ref{3.10}) uniquely determines
our solution and its properties ($a_1$, $a_2$ and $R$) for any
$\vepps$.  This solution is shown in Fig.1 (function A). This
solution looks like a shell with radius $R$.

Using (\ref{3.8}), one obtains the following expression for energy of
such gluon cluster
\beq
E^\vepps = \frac{1}{\vepps} \, E^{\vepps=1} = M \, E^{\vepps=1},
\label{3.11}
\eeq
where $E^{\vepps=1}=110.75$ is the energy of field configuration if
$\vepps=1$. Expression (\ref{3.11}) is intuitively clear. Indeed,
the pure Yang-Mills theory is scale invariant and has no
mass-dimensional parameters. Modified Yang-Mills theory (\ref{2.1})
has such parameter $\vepps=1/M$ and the mass of gluon objects under
investigation is proportional to this parameter.

Now to predict the physical mass $M_{cluster}$ and effective radius
we should have a prediction of the value of parameter $\vepps=1/M$.
In this paper, following \cc{e4}, we proposed that $M \simeq 0.59 \pi
GeV $, and our model gives the following prediction of the  mass and
effective radius of investigated gluon clusters:
\beq
M=1/\vepps \simeq 0.59 \pi GeV,  \qquad
M_{cluster} \simeq 205 \, GeV, \qquad
R \simeq 0.15 fm \label{3.12}
\eeq

In the next section we give some conclusions and perspectives of
such investigations are discussed.

\section{Conclusions.}

The aim of this paper is to show that quantum fluctuations of
nonabelian Yang-Mills field can lead to generation of the stable
cluster finite-energy solution. In our work we used the
gauge-invariant approach \cc{e1,e2,e3,ale1,ale2} in  which such
quantum fluctuation  should be taken into account by adding
high-derivative terms to the pure Yang-Mills lagrangian.  In the
present work, we investigated the effective $SU(2)$ Yang-Mills
theory and chromomagnetic spherically symmetrical field
configurations.

One of the interesting consequences of this effective theory is a
fact that for the investigated field configuration there exists a
critical value of energy density. This fact is due to the physical
condition of continuity of energy density. Such condition allowed
us to construct the cluster solution for all space points. We
predicted that the mass of such object should be about two hundred
GeV and effective radius should be about $0.2$ fm.

Of course, we do not give a comprehensive investigation of  this
effective Yang-Mills theory. The questions about dyon solution or
about a role of contributions from other high derivative modified
term in pure Yang-Mills lagrangian are clear now. But maybe the most
important question in such investigation is about physical
consequences of existence of such gluon cluster objects and about
their experimental status. All of these questions should be the
themes for  future investigation.

\section*{Acknowledgments.}

The author is grateful to Dr.~A.N.~Sobolevski\u{\i} and
Dr.~I.O.~Cherednikov for useful discussions. This work was
supported in part by RFBR under Grant No.96-15-96674.

\end{document}